\documentclass[12pt]{article}
\usepackage{a4wide}

\begin{document}
\author{Bernard Linet \thanks{E-mail: linet@lmpt.univ-tours.fr}\\
\small Laboratoire de Math\'ematiques et Physique Th\'eorique \\
\small CNRS/UMR 6083, F\'ed\'eration Denis Poisson \\
\small Universit\'e Fran\c{c}ois Rabelais, 37200 TOURS, France}

\title{\bf Bending of a light ray within a dispersive medium in a uniform gravitational field}

\date{}
\maketitle
\thispagestyle{empty}

\begin{abstract}
Recently, Dressel {\em et al} determined the vertical deflection of a light ray in a medium,
with strong frequency-dependent dispersion, at rest in a uniform gravitational field. We take up this question 
within the general relativistic theory of the propagation of light in a dispersive medium, due to Synge.
\end{abstract}

In a recent paper \cite{dressel}, Dressel {\em et al} determined the vertical deflection $\Delta z$ of a light
ray in a medium, with strong
frequency-dependent dispersion, at rest in a uniform gravitational field. Following their method, they found
\begin{equation}\label{formule}
\Delta z\approx -\frac{gL^2v_p}{2c^2v_g} ,
\end{equation}
where $g$ is the acceleration due to the Earth gravity, $L$ the covered distance and $v_p$ and $v_g$ are
respectively the phase velocity and the group velocity of this dispersive medium.

We take up this question in a more general approach. We consider the general relativistic theory of
the propagation of light in a dispersive medium, proposed by Synge \cite{synge} a long time ago. We first summarize
briefly 
this theory in a curved space-time. Then, we carry out the exact calculations in the case of the Rindler space-time
representing a uniform gravitational field. For a weak acceleration of gravity, we find again
formula (\ref{formule}). In our knowledge, this approach of the problem has not been already performed.

We consider a space-time which is described by a metric $g_{\mu \nu}$ in a coordinate system $(x^{\mu})$.
We adopt the signature $(-+++)$ for the metric. We use units in which $c=1$. The dispersive medium is
described by its 4-velocity $V^{\mu}$, satisfying $g_{\mu \nu}V^{\mu}V^{\nu}=-1$.
We now consider the propagation of a wave in a dispersive medium. It is defined by the equation 
$S(x^{\mu})= {\rm constant}$ where $S$ is a function. The vector of propagation $k^{\mu}$ is defined by
\begin{equation}\label{kmu}
k_{\mu}=\partial_{\mu}S .
\end{equation}
The frequency $\omega$ of the wave, relative to $V^{\mu}$, has the expression
\begin{equation}\label{omega}
\omega = -k_{\mu}V^{\mu} .
\end{equation}
The phase velocity $v_p$, relative to $V^{\mu}$, is given by
\begin{equation}\label{defvp}
\frac{1}{v_{p}^{2}}=1+\frac{g^{\mu \nu}k_{\mu}k_{\nu}}{\left( k_{\mu}V^{\mu}\right)^2} .
\end{equation}

In classical theory of optics, a dispersive medium is characterized by the refractive index $n$ which is the reciprocal
of the phase velocity, i.e.
\begin{equation}\label{n}
n=\frac{1}{v_p} ,
\end{equation}
and consequently the relation of dispersion takes the form
\begin{equation}\label{dispersion}
g^{\mu \nu}k_{\mu}k_{\nu}-\left( n^2-1\right) \omega^2=0 .
\end{equation}
The refractive index $n$ is a given function $n(x^{\mu},\omega )$ which characterizes the dispersive medium.

The equations of the light rays are derived from the Hamiltonian function
\begin{equation}\label{h}
{\cal H}\left( x^{\mu},k_{\nu}\right) =\frac{1}{2}\left[ g^{\mu \nu}( x^{\lambda}) k_{\mu}k_{\nu}
-\left( n^2( x^{\lambda},\omega )-1\right) \omega^2 \right]
\end{equation}
where $\omega =\omega ( x^{\lambda},k_{\sigma})$ given by (\ref{omega}). The characteristic curves of
${\cal H}(x^{\mu},k_{\nu})=0$ are the light rays. From the usual canonical equations, we find thus the 
following system of differential equations for a parameter $\lambda$: 
\begin{equation}\label{dxmu}
\frac{dx^{\mu}}{d\lambda}=g^{\mu \nu}k_{\nu}+\left[ \left( n^2-1\right) \omega +
n\frac{\partial n}{\partial \omega}\omega^2\right] V^{\mu} ,
\end{equation}
\begin{equation}\label{dkmu}
\frac{dk_{\mu}}{d\lambda}=-\frac{1}{2}\partial_{\mu}g^{\alpha \beta}\, k_{\alpha}k_{\beta}+
n\partial_{\mu}n \, \omega^2-\left[ \left( n^2-1\right) \omega +n\frac{\partial n}{\partial \omega}\omega^2\right]
k_{\nu}\partial_{\mu}V^{\nu} ,
\end{equation}
and we have to add constraint (\ref{dispersion}). 

The curves $x^{\mu}(\lambda )$ define the congruence of
world-lines of light rays, i.e. the trajectories of the pulses of light. So, the vector $dx^{\mu}/d\lambda$ 
must be timelike or null. The group velocity $v_g$, relative to $V^{\mu}$, is given by
\begin{equation}\label{defvg}
v_{g}^{2}=1+\frac{\displaystyle g_{\mu \nu}\frac{dx^{\mu}}{d\lambda}\frac{dx^{\nu}}{d\lambda}}
{\displaystyle \left( \frac{dx^{\mu}}{d\lambda}V_{\mu}\right)^2} .
\end{equation}
According to (\ref{dxmu}), we easily see that
\begin{equation}\label{rel}
g_{\mu \nu}\frac{dx^{\mu}}{d\lambda}\frac{dx^{\nu}}{d\lambda}=n^2\omega^2
\left[ 1-\left( \frac{\partial}{\partial \omega}(n\omega )\right)^2\right] \quad {\rm and}\quad
\frac{dx^{\mu}}{d\lambda}V_{\mu}=-n\omega \frac{\partial}{\partial \omega}(n\omega ) .
\end{equation}
By inserting expressions (\ref{rel}) into (\ref{defvg}), we obtain \footnote{As noticed by Bi\v{c}\'ak and Hadrava
\cite{bicak}, there is a misprint in the book of Synge \cite{synge}.}
\begin{equation}\label{vg}
v_g=\frac{1}{\displaystyle \frac{\partial}{\partial \omega}(n\omega )} .
\end{equation}
We assume that the solutions to differential equations (\ref{dxmu}) and (\ref{dkmu}) yield $v_g\leq 1$.

We add the following property in the case where there exists a vector $\xi^{\mu}$ which is
a Killing vector for the metric and also for the dispersive medium, i.e.
\begin{equation}\label{xi}
{\cal L}_{\xi}g_{\mu \nu}=0 \quad , \quad {\cal L}_{\xi}n=0 \quad {\rm and}\quad {\cal L}_{\xi}V^{\mu}=0 
\end{equation}
where ${\cal L_{\xi}}$ is the Lie derivative with respect to $\xi^{\mu}$. 
From equations (\ref{dxmu}) and (\ref{dkmu}), we can straightforwardly deduce the following 
differential equation along the light rays:
\begin{equation}\label{constante}
\frac{d}{d\lambda}\left( \xi^{\mu}k_{\mu}\right) = 0 ,
\end{equation}
and hence we get that $\xi^{\mu}k_{\mu}$ is a constant of motion.

We are now in a position to consider the case of the Rindler space-time described by the metric
\begin{equation}\label{rindler}
ds^2=-\left( 1+gz\right)^2\left( dx^0\right)^2+dx^2+dy^2+dz^2 ,
\end{equation}
well defined for $gz>-1$. When $|  gz | \ll 1$, metric (\ref{rindler}) represents a Newtonian potential $U=-gz$
in Galilean coordinates where $g$ is the acceleration of the gravity. We assume that the dispersive medium
is at rest in these coordinates and moreover that the refractive index $n$ does not depend on the
position, $n=n(\omega )$. The 4-velocity $V^{\mu}$ has the following components:
\begin{equation}\label{vmu}
V^0=\frac{1}{1+gz} \quad {\rm and}\quad V^x=V^y=V^z=0 .
\end{equation}
In metric (\ref{rindler}) with 4-velocity (\ref{vmu}), there are three Killing vectors $\partial /\partial x^0$, 
$\partial /\partial x$ and $\partial /\partial y$.

We consider a light ray in the plane $x=0$, horizontally emitted at the point $y=0$ and $z=0$. The constants of
motion (\ref{constante}) are denoted
\begin{equation}\label{k0}
k_0=-\omega_0 \quad , \quad k_x=0\quad {\rm and} \quad k_y=\omega_0k
\end{equation}
where $\omega_0$ and $k$ are constants. According to (\ref{omega}), we notice that
\begin{equation}\label{omega0}
\omega =\frac{\omega_0}{1+gz} .
\end{equation}
We see from (\ref{omega0}) that $\omega_0$ is the frequency of the wave, relative to $V^{\mu}$, at $z=0$.
We write down equation (\ref{dxmu}) in this particular case
\begin{equation}\label{dxmu0}
\frac{dx}{d\lambda}=0 \quad , \quad \frac{dy}{d\lambda}=\omega_0k \quad {\rm and} \quad \frac{dz}{d\lambda}=k_z .
\end{equation}
We now give the relation of dispersion (\ref{dispersion}) with expressions (\ref{k0}) and (\ref{omega0})
\begin{equation}\label{e1}
\omega_{0}^{2}k^2+\left( k_z\right)^2-n^2(\omega )\frac{\omega_{0}^{2}}{\left( 1+gz\right)^2} =0.
\end{equation}
At the point $y=0$ and $z=0$, we have $k_z=0$ since the light is horizontally emitted. Consequently, we get 
from relation (\ref{e1}) at $z=0$ the value of the constant $k$
\begin{equation}\label{e2}
k=n(\omega_0) .
\end{equation}
Finally, we obtain the desired expression of $k_z$
\begin{equation}\label{e3}
\left( k_z\right)^2=\frac{n^2(\omega )\omega_{0}^{2}}{\left( 1+gz\right)^2}-n^2(\omega_0)\omega_{0}^{2} .
\end{equation}
Taking into account (\ref{dxmu0}), we find from (\ref{e3}) the exact differential equation
\begin{equation}\label{exact}
\left( \frac{dz}{dy}\right)^2=\frac{\displaystyle n^2\left( \frac{\omega_0}{1+gz}\right)}
{\left( 1+gz\right)^2n^2(\omega_0)} -1
\end{equation}
giving the vertical deflection of the light ray when the refractive index $n(\omega )$ of the
dispersive medium is known.

We are now ready to determine the vertical deflection in the limit where $| gz | \ll 1$. We expand expression 
(\ref{exact}) linearly in $gz$
\begin{equation}\label{f1}
\left( \frac{dz}{dy}\right)^2\approx -\frac{\displaystyle 2\frac{dn}{d\omega}(\omega_0)\, \omega_0gz}{n(\omega_0)} -2gz .
\end{equation}
Making use of (\ref{defvp}) and (\ref{vg}) expressed at $z=0$, we rewrite (\ref{f1}) under the form
\begin{equation}\label{f2}
\left( \frac{dz}{dy}\right)^2\approx -2\left( \frac{v_p}{v_g}\right) gz ,
\end{equation}
with $z<0$ of course. Differential equation (\ref{f2}) leads obviously to formula (\ref{formule}) of Dressel
{\em et al} \cite{dressel}.

We have obtained a general result. We have not made the slow light approximation that the refreactive index is
linear in the frequency.


\begin{thebibliography}{99}
\bibitem{dressel} J. Dressel, S. G. Rajeev, J. C. Howell and A. N. Jordan, Phys. Rev. A {\bf 79} 013834 (2009).  
\bibitem{synge} J. L. Synge, {\em Relativity: The General Theory}, Chap. XI, North-Holland Publishing 
Company (1966).
\bibitem{bicak}J. Bi\v{c}\'{a}k and P. Hadrava, Astron. Astrophys.  {\bf 44} 389 (1975).
\end{thebibliography}
\end{document}